\documentclass[iop,apjl]{emulateapj}

\newcommand{\simless}{\mathbin{\lower 3pt\hbox
      {$\rlap{\raise 5pt\hbox{$\char'074$}}\mathchar"7218$}}} 
\newcommand{\simgreat}{\mathbin{\lower 3pt\hbox
     {$\rlap{\raise 5pt\hbox{$\char'076$}}\mathchar"7218$}}} 

\slugcomment{\today}

\shorttitle{ATCA observations of debris disks}
\shortauthors{Ricci et al.}

\begin{document}


\title{An ATCA survey of debris disks at 7 millimeters}


\author{L. Ricci}
\affil{Harvard-Smithsonian Center for Astrophysics, 60 Garden Street, Cambridge, MA 02138, USA}

\and

\author{S. T. Maddison}
\affil{Centre for Astrophysics \& Supercomputing, Swinburne University, PO Box 218, Hawthorn, VIC 3122, Australia}

\and

\author{D. Wilner, M. A. MacGregor}
\affil{Harvard-Smithsonian Center for Astrophysics, 60 Garden Street, Cambridge, MA 02138, USA}

\and

\author{C. Ubach}
\affil{National Radio Astronomy Observatory, North American ALMA Science Center,
520 Edgemond Road, Charlottesville, VA 22903, USA}

\and

\author{J. M. Carpenter}
\affil{Department of Astronomy, California Institute of Technology, MC 249-17, Pasadena, CA 91125, USA}

\and

\author{L. Testi}
\affil{European Southern Observatory (ESO) Headquarters, Karl-Schwarzschild-Str. 2, D-85748 Garching bei Muenchen, Germany}

\email{luca.ricci@cfa.harvard.edu}


\begin{abstract}

We present ATCA continuum observations at a wavelength of 6.8 mm of five debris disks: $\beta$ Pictoris, q$^1$ Eridani, HD 107146, HD 181327, and HD 95086. These observations provide the detection at the longest wavelengths obtained to date for all these debris disks. By combining our 6.8 mm data with previous detections at shorter sub-millimeter/millimeter wavelengths we measure the long wavelength spectral index of these sources. We then use previous estimates for the temperature of the emitting dust to derive the spectral index of the dust emissivity. Under the assumption that all the detected flux comes from dust only, we constrain the slope of the solid size distribution, assumed to be a power-law.
The values that we infer for the slope of the size distribution range between about 3.36 and 3.50.
We compare our findings with the case of the Fomalhaut debris disk and use these results to test the predictions of collisional cascades of planetesimal belts. 

\end{abstract}



\section{Introduction}
\label{sec:intro}


Observations of debris disks made of cold dust around nearby stars provide crucial information to the process of planet formation  \citep[e.g.][]{Zuckerman:2001,Matthews:2014}. The dust grains observed in these systems are thought to be produced by collisions between km-sized planetesimals (comets and asteroids), leftovers of the planetary formation process \citep[see][]{Wyatt:2008}. The highly destructive mutual encounters between planetesimals which generate the observed dust are probably triggered by the dynamical interaction with one or more planets in the system \citep{Mustill:2009}.

Because of their extremely faint emission, km-sized planetesimals cannot be directly observed outside the Solar System. However, information on their physical and dynamical properties can be extracted from the size distribution of the emitting smaller dust grains.
Observations in the sub-millimeter/millimeter can quantify the departure of the spectrum from the Rayleigh-Jeans regime ($\propto \nu^2$) which is due to a combination of temperature and emissivity effects. As the temperature of dust grains can be inferred from observations in the infrared, the frequency-dependence of the dust emissivity at sub-mm/mm can be estimated, and this is related to the size distribution of $\sim 1-10$~mm-sized dust grains present in a debris disk \citep{Draine:2006,Ricci:2012,Gaspar:2012}. Observations at long wavelengths have the advantage of minimizing the departures from the Rayleigh-Jeans approximation as well as the effect of optical depth because of the lower dust emissivities \citep[e.g.][]{Testi:2014}.

This method can be used to test collisional models of debris disks which provide different predictions for the solid size distribution depending on the dynamical state \citep[e.g. velocity dispersion,][]{Pan:2012} or physical conditions \citep[e.g. tensile strength,][]{Durda:1997,Gaspar:2012} of the large colliding bodies. The reference model is the collisional cascade of large bodies at the steady state, which predicts a power-law solid size distribution ($n(a) \propto a^{-q}$) with exponent $q = 3.51$, as a result of the collective dynamical interaction of particles caused by inelastic collisions and fragmentation \citep{Dohnanyi:1969}. 
This standard model assumes a single constant tensile strength\footnote{The tensile strength of a body is defined as the minimum energy per unit mass required to disrupt it catastrophically.} and velocity dispersion for all bodies regardless of size.
More recent numerical simulations and analytical calculations have relaxed these assumptions and derived steeper grain size distributions ($q \simgreat 3.5 - 3.6$) in the case in which either the tensile strength or the velocity dispersion of the large bodies are steep functions of their size \citep{Pan:2012, Gaspar:2012}.

We present the results from observations with the Australia Telescope Compact Array\footnote{The Australia Telescope Compact Array is part of the Australia Telescope National Facility which is funded by the Commonwealth of Australia for operation as a National Facility managed by CSIRO.} (ATCA) of five debris disks at 6.8~mm: $q^1$ Eri, HD 107146, HD 181327, HD 95086, $\beta$ Pic. In the case of $\beta$ Pic, the ATCA observations were already introduced in \citet{Wilson:2011}.  
We combine these data with previous observations at shorter wavelengths to estimate the $q$-values of the slope of the size distribution for each debris disk in our sample, and compare our results with the predictions of the collisional cascade models. 

In Section~\ref{sec:sample} we present the main properties of the debris disks in our sample. Sections~\ref{sec:obs} and~\ref{sec:results} describe the ATCA observations and their results, respectively. Results on the slope of the solid size distribution are presented and discussed in Sections~\ref{sec:slope} and~\ref{sec:discussion}. Section~\ref{sec:summary} summarizes the main findings of this study.

\section{The sample}
\label{sec:sample}

In this section we present the main properties of the five debris disks in our sample. These sources were selected to be 1) at declinations lower than 20 degrees in order to be observable with the ATCA array, 2) relatively bright sources in the sub-mm, i.e. with a flux density at 0.85~mm greater than 30 mJy, and 3) relatively compact, i.e. with angular size lower than $15''$ to maximize surface brightness sensitivity.  

\subsection{q$^{1}$ Eri}

q$^{1}$ Eridani (HD 10647) is a F9 main sequence star at a distance of 17.4~pc. Its estimated age is $\approx 1.7 - 3.2$ Gyr \citep{Moro-Martin:2015, Bonfanti:2015}. q$^{1}$ Eri hosts a $\sim 1~M_{\rm{Jup}}$ planet which was discovered with the radial velocity technique at a distance of $\approx 2$ AU from the star \citep[see][]{Butler:2006}. 

Evidence for a dusty belt has been obtained from observations in the infrared~\citep{Zuckerman:2004, Chen:2006, Trilling:2008} and sub-millimeter~\citep{Liseau:2008}.  Recently, \citet{Moro-Martin:2015} have fitted the disk Spectral Energy Distribution (SED) including far-IR fluxes with the \textit{Herschel Space Observatory}. They inferred a dust temperature of $\approx 49$ K at a stellocentric distance of about 40 AU. 

\subsection{HD 107146}

HD 107146 is a Solar-like G2 type star. At a distance of 27.5~pc and with an age of $\sim 80 - 200$ Myr, this is the brightest known debris disk at infrared wavelengths around a G-type star \citep{Silverstone:2000,Moor:2006}. The disk was spatially resolved both in scattered light with the Hubble Space Telescope \citep{Ardila:2004,Ertel:2011,Schneider:2014} and in dust thermal emission with sub-millimeter single dish \citep{Williams:2004} and interferometric observations \citep{Corder:2009,Hughes:2011,Ricci:2015}.

Spitzer spectroscopic and photometric data revealed the presence of warm dust ($\sim 120$~K) located $\sim 5 - 15$ AU from the star \citep{Morales:2011}. A colder dust component extending between stellocentric radii of about 30 and 150 AU dominates the sub-mm emission \citep{Ricci:2015}.
Recent observations with the Atacama Large Millimeter Array (ALMA) show a decrease in the surface brightness of the disk at $\approx 80$ AU from the star, which could be due to dynamical interaction with a low-mass terrestrial planet in the system \citep{Ricci:2015,Kenyon:2015}.

%


\subsection{HD 181327}

HD181327 is a F6-type star at a distance of $\approx 51$ pc \citep{Zuckerman:2004,Casagrande:2011}. It is located in the $\beta$ Pic moving group with an age of 23 $\pm$ 3 Myr \citep{Mamajek:2014}.  

In order to reproduce the emission in the infrared, \citet{Mittal:2015} invoked a warm dust component at a stellocentric distance of $\approx 0.3 - 2$ AU (temperature of $\approx 340$ K), as well as a colder component with $\approx 82$ K further from the star. \citet{Schneider:2006,Schneider:2014} presented spatially resolved observations of the debris disk in scattered light at optical and near-infrared wavelengths with the Hubble Space Telescope. These observations show that a relatively narrow ring at a distance of $\approx 88$ AU from the star accounts for most of the total flux from the disk.

\subsection{HD 95086}

HD 95086 is an A8-type member of the Lower Centaurus Crux Association at a distance of $\approx 90$ pc \citep{deZeeuw:1999,vanLeeuwen:2007} and age of $17 \pm 6$ Myr. A planet with a mass of $5 \pm 2~M_{Jup}$ has been directly imaged at a projected distance of 56 AU from the star \citep{Rameau:2013}.  

The observed disk SED of HD 95086 is best described with two dust temperatures: a warm component
at $\approx$ 175 K and a colder component at $\approx$ 55 K with an outer radius of $\sim 200$ AU \citep{Su:2015}. 
The analysis of spatially resolved images in the far-infrared indicates the presence of dust grains at further distances, possibly a diffuse halo of small particles thrown into highly eccentric or hyperbolic orbits by stellar radiation pressure. 

\subsection{$\beta$ Pic}

$\beta$ Pictoris (HD 39060) is an A6 type star at a distance of 19.4 pc and with an age of 23 $\pm$ 3 Myr \citep{Gray:2006,vanLeeuwen:2007,Mamajek:2014}.

The star is surrounded by the first debris disk to be imaged in scattered light \citep{Smith:1984}. Subsequent observations have shown the existence of a massive planet at $\sim 10$ AU from the star \citep{Lagrange:2010}, comets falling toward the star within a few AU \citep{Vidal-Madjar:1994}, atomic gas extending out to $\sim$ 300 AU, sub-mm dust emission out to $\approx 100$ AU \citep{Wilner:2011}, as well as a clear azimuthal asymmetry in the spatial distribution of molecular gas \citep{Dent:2014}. 

\section{ATCA Observations and data reduction}
\label{sec:obs}

ATCA observations of the 6.8~mm dust continuum toward the five disks in our sample
were obtained using the Compact Array Broadband Backend (CABB) digital filter
bank \citep{Wilson:2011}. 

Observations were carried out on September 14 2009 for $\beta$ Pic \citep[see][for a presentation of these observations]{Wilson:2011}, and between July 27 and August 5 2012 for the other four disks. The ATCA array was in the H214 configuration for the observations of $\beta$ Pic (baselines between 82 and 247 m), and in the H75 configuration for the other observations (baselines between 31 and 89 m). The array comprises six 22m diameter antennas, but data from the stationary sixth antenna, located about 6 km from the others, was discarded because of the higher phase noise on the much longer baselines. To obtain the best sensitivity in the continuum the correlator was set to cover the full 4-GHz CABB effective bandwidth. The mean frequency of the observations is 44.0 GHz, corresponding to a wavelength of about 6.8~mm.

\begin{deluxetable*}{lcccc}
\tablecaption{ATCA Observations\label{tbl:obs}}
\tablehead{
  \colhead{Source}  &
  \colhead{R.A.}     &
  \colhead{Dec.} &
  \colhead{On-source time} &
  \colhead{Gain calibrator(s)} \\
  \colhead{}  &
  \colhead{}    &
  \colhead{} &
  \colhead{[minutes]} &
  \colhead{} 
}
\startdata

q$^1$ Eri  &  01:42:29.32 & $-$53:44:27.0  & 488 & 0208-512 \vspace{1mm} \\
HD 107146  &  12:19:06.50 & +16:32:53.9  & 490 & 1236+077 \vspace{1mm} \\
HD 181327  &  19:22:58.94 & $-$54 32 17.0  & 721 & 1933-400, 1941-554 \vspace{1mm} \\
HD 95086  &  10:57:03.02 & $-$68:40:02.5  & 712 & J1147-6753 \vspace{1mm} \\
$\beta$ Pic  &  05:47:17.09 & $-$51:03:59.4  & 680 & 0537-441 \\

\enddata

\end{deluxetable*}

Table~\ref{tbl:obs} summarizes the coordinates of each target, the on-source times and gain calibrators.
We used 0537-441, 1253-055 and 1921-293 as bandpass calibrators, whereas Uranus, Neptune and 1934-638 were used to calibrate the absolute flux scale of the data.  
We adopt the typical value of $10\%$ for the uncertainty on the absolute flux scale for ATCA observations at this wavelength\citep[e.g.][]{Ubach:2012}. This uncertainty is in agreement with the variation of the flux for the gain calibrators as measured in different days during our observations.

The ATCA data were calibrated using the \texttt{miriad} software package. Imaging and deconvolution were performed with the standard \texttt{miriad} routines \texttt{invert}, \texttt{clean} and \texttt{restore}.

\section{Results of the ATCA observations}
\label{sec:results}

Continuum emission from all the five disks was detected with our ATCA observations at 6.8~mm (Table~\ref{tbl:obs_results}). 
The achieved signal-to-noise ratios range between $\approx 4$ (HD95086) and $\approx 12$ (HD 181327). 

As expected, the emission is spatially unresolved for the four disks observed in the H75 array configuration (q$^1$ Eri, HD107146, HD181327, HD95086), as the angular resolution for these observations is significantly larger than the angular extent of those sources. For these four spatially unresolved disks, the flux density was extracted using the \texttt{uvfit} routine that fits the calibrated visibilities with a point-source model.

\begin{figure}[t!]
\centering
\includegraphics[scale=0.5]{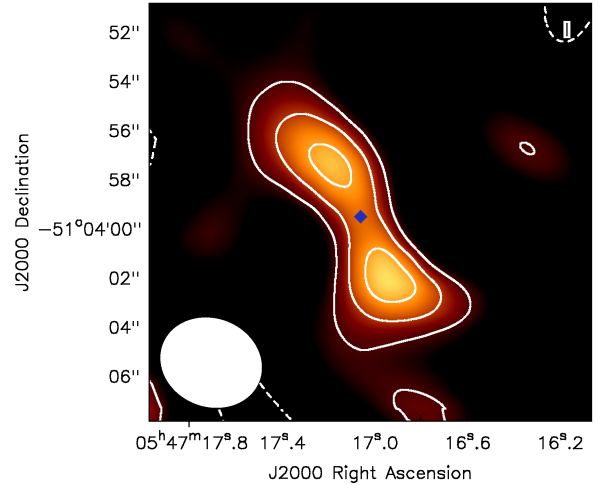}   
\vspace{-10mm}
\caption{ATCA 6.8 mm continuum map of $\beta$ Pic. Contours are drawn at $-2\sigma$ (dashed lines), 2$\sigma$, 3$\sigma$, 4$\sigma$ (solid), where 1$\sigma =$ 23.0 $\mu$Jy/beam is the rms noise of the map. The blue symbol toward the center of the map shows the location of the star. The white ellipse in the lower left corner represents the synthesized beam with FWHM size of $4.18'' \times 3.56''$ and a position angle of 68.7 deg.}
              \label{fig:betapic}
\end{figure}

In the case of $\beta$ Pic, the higher angular resolution achieved with the H214 array configuration allowed us to spatially resolve the disk (Fig.~\ref{fig:betapic}). The disk is elongated along the northeast-southwest direction, as seen in all the previous maps for this source at shorter wavelengths.  The map at 6.8~mm also shows higher surface brightness toward the southwest lobe of the disk, as previously seen at 0.87~mm \citep{Dent:2014} and 1.3~mm~\citep{Wilner:2011} with better signal-to-noise.
For this disk, the total flux density reported in Table~\ref{tbl:obs_results} was obtained by integrating the surface brightness over the area showing emission at $\simgreat~2\sigma$ above the background level in the map. 

\begin{deluxetable*}{lccccc}
\tablecaption{Results of the ATCA Observations\label{tbl:obs_results}}
\tablehead{
  \colhead{Source}  &
  \colhead{Beam Size}     &
  \colhead{Beam P. A.$^{a}$} &
  \colhead{Flux density at 6.8mm} &
  \colhead{RMS noise} &
  \colhead{Total uncertainty of flux density$^{b}$} \\
  \colhead{}  &
  \colhead{}  &
  \colhead{[degrees]} &
  \colhead{[$\mu$Jy]} &
  \colhead{[$\mu$Jy/beam]} &
  \colhead{[$\mu$Jy]}
}
\startdata

q$^1$ Eri  &  $14.81'' \times 10.32''$ &  278.4 & 92.6 & 13.8   &  16.6 \vspace{1mm} \\
HD 107146  & $12.37'' \times 10.77''$ & 0.1 & 166.0 & 19.0 &  25.2  \vspace{1mm} \\
HD 181327  & $13.24'' \times 9.82''$ & 277.4 & 145.0 & 12.6 &  19.2    \vspace{1mm} \\
HD 95086  & $12.68'' \times 11.04''$  & 284.9  & 61.9 &   14.6 &  15.9 \vspace{1mm} \\
$\beta$ Pic       &  $4.18'' \times 3.56''$ & 68.7    &     240.0$^{c}$     &    23.0        &      33.2     \\
\enddata
\tablenotetext{a}{The beam Position Angle is defined as the angle east of north to the disk major axis.}
\tablenotetext{c}{The total uncertainty of flux density includes the RMS noise and the uncertainty on the calibration of the absolute flux scale.}
\tablenotetext{c}{The observations of $\beta$ Pic are affected by spatial filtering so the flux quoted here has to be interpreted as a lower limit of the total flux density from the source.}

\end{deluxetable*} 

\section{The slope of the solid size distribution} 
\label{sec:slope}

Models of collisional cascades for planetesimal belts predict size distributions which are well approximated by power-laws $n(a) \propto a^{q}$, typically with $3 < q < 4$, and between blow-out grain sizes (typically of the order of $\sim \mu$m) and the sizes of the solids which initiate the cascade, i.e. $\sim 1 - 100$ km~\citep[e.g.][]{Dohnanyi:1969}. 

Under these conditions, \citet{Draine:2006} derived a relation between the spectral index $\beta$ of the dust emissivity ($\kappa_{\nu} \propto \nu^{\beta}$), the spectral index $\beta_s$ of small particles only (i.e. with sizes much smaller than the observing wavelengths), and the power-law index $q$ of the solid size distribution. This formula reads: $\beta = \beta_s (q - 3)$.


Since the dust thermal emission in debris disks is optically thin, the flux density $F_{\nu} \propto B_{\nu}(T_{\rm{dust}})\kappa_{\nu}\propto B_{\nu}(T_{\rm{dust}})\nu^{\beta}$, where $B_{\nu}(T_{\rm{dust}})$ is the Planck function evaluated at the temperature of the dust in the disk.
For debris disks around stars with spectral types of G and earlier, such as in our sample, the dust is usually warm enough ($k_{B}T_{\rm{dust}} >> h \nu$) for the Planck function to be well approximated by the Rayleigh-Jeans form ($B_{\nu}(T_{\rm{dust}}) \propto \nu^2$). A more accurate approximation is obtained considering the Planck function as a power-law, $B_{\nu}(T_{\rm{dust}}) \propto \nu^{\alpha_{\rm{Planck}}}$ with $\alpha_{\rm{Planck}} = \alpha_{\rm{Planck}}(T_{\rm{dust}}) \simless 2$, and equal to 2 in the Rayleigh-Jeans regime. 

The flux density of debris disks at sub-mm and mm wavelengths is a power-law in frequency ($F_{\nu} \propto \nu^{\alpha_{\rm{mm}}}$), and Draine's formula can be written in terms of $q$ as:

\begin{equation} \label{eq_draine}
q \approx \frac{\alpha_{\rm{mm}}-\alpha_{\rm{Planck}}}{\beta_s} + 3,
\end{equation}
\citep[see][]{Ricci:2012}.

For $\beta_s$ we consider a value of $1.8 \pm 0.2$, consistent with observations of small dust grains in several diffuse interstellar clouds, molecular clouds, as well as with predictions for small grains with chemical compositions expected in protoplanetary disks \citep[e.g.,][and references therein]{Draine:2006,Testi:2014}. Hence, by measuring the sub-mm/mm spectral index $\alpha_{\rm{mm}}$ and by inferring $\alpha_{\rm{Planck}}$ from the dust temperature estimated for each disk, we can use Eq.~\ref{eq_draine} to infer the slope $q$ of the solid size distribution\footnote{Here we are implicitly assuming that all the emission observed at sub-mm/mm wavelengths comes from dust. The validity of this assumption will be discussed at the end of Section~\ref{sec:discussion}.}.

Table~\ref{tbl:results} lists the flux density of each disk at a sub-mm/mm wavelength $\lambda_{1}$ reported in the literature which is combined with our ATCA flux density at 6.8~mm to measure $\alpha_{\rm{mm}}$ ($= \log(F_{\nu,\lambda_{1}}/F_{\nu,\lambda=6.8\rm{mm}})/\log(\lambda_{1}/6.8\rm{mm})$ ), the dust temperature $T_{\rm{dust}}$, the spectral index of the Planck function $\alpha_{\rm{Planck}}$  ($= \log(B_{\nu,\lambda_{1}}/B_{\nu,\lambda=6.8\rm{mm}})/\log(\lambda_{1}/6.8\rm{mm})$ ), and the slope of the solid size distribution $q$ as derived using Eq.~\ref{eq_draine}. 
For HD 107146 and $\beta$ Pic we chose the accurate flux measurements extracted from published ALMA observations. For the other three disks for which ALMA observations have not been obtained yet, we chose flux measurements with Herschel ($q^1$ Eri, HD 95086) and APEX (HD 181327), see references in Table~\ref{tbl:results}. In the cases in which Herschel and APEX observations are both present, we have verified that the $q$-values obtained using either one or the other flux density are consistent with each other at the $\simless~2\sigma$ level. 

For $\beta$ Pic, the sub-mm/mm flux reported in the table was extracted from ALMA observations at 0.87 mm \citep{Dent:2014}. Like our ATCA observations for the same source, also the ALMA observations were affected by spatial filtering of the emission on the largest scales. The shortest projected baselines of the ALMA have $uv$-distances of 17 k$\lambda$. In order to compare emission on the same spatial scales, we derived the emission including only visibilities associated to $uv$-distances of 17 k$\lambda$ and larger from our ATCA dataset. We derived a value of 185 $\pm$ 33.6 $\mu$Jy  (rms noise on the map is 28.0 $\mu$Jy/beam). We used this value to derive the spectral index $\alpha_{\rm{mm}}$, and associated uncertainty, reported in Table~\ref{tbl:results}.

\begin{figure*}[t!]
\centering
\includegraphics[scale=0.5]{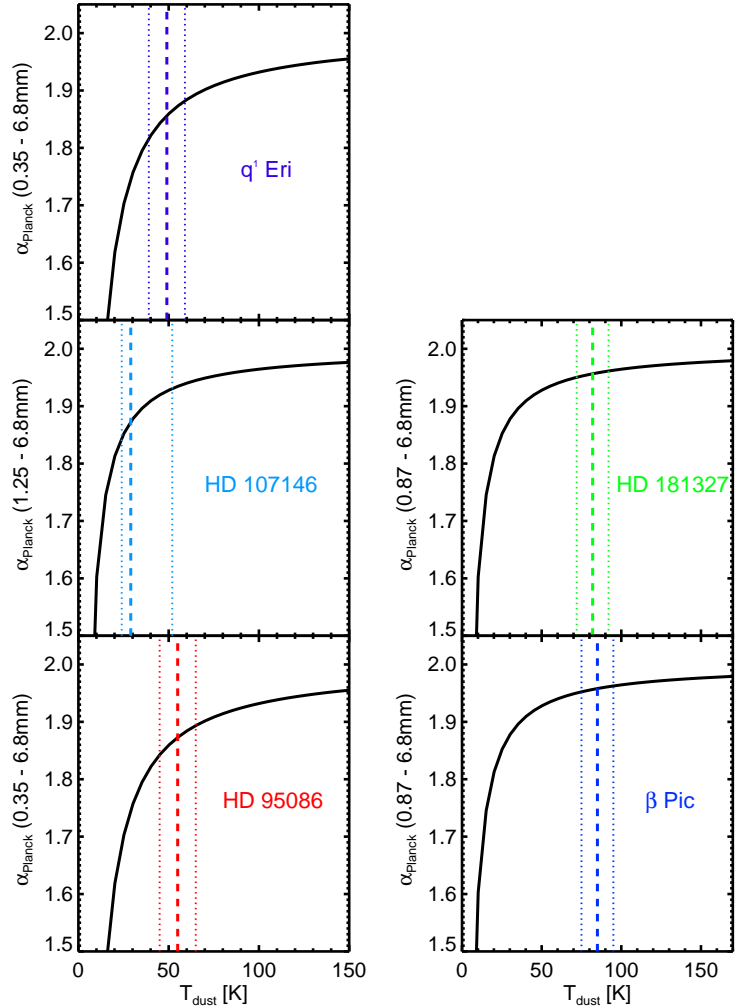}   
\vspace{-3mm}
\caption{Spectral index of the Planck function as a function of dust temperature. In each panel, the dashed vertical line shows the dust temperature adopted in our analysis, and the dotted vertical lines show the limiting values of the intervals reported in Table~\ref{tbl:results}.}
              \label{fig:dust_temperature}
\end{figure*}

In all cases but HD 107146, the dust temperature $T_{\rm{dust}}$ was inferred from fitting of the SED at infrared and sub-mm wavelengths. A reasonable uncertainty of $\pm$ 10 K was considered to derive an uncertainty on $\alpha_{\rm{Planck}}$. In the case of HD 107146, we considered the interval of temperatures for dust with size of $\approx 1$ mm at the stellocentric radii of $\approx 30 - 150$ AU, as estimated by \citet{Ricci:2015}. Note that this interval includes the estimate of $\approx 51$ K as derived through SED fitting by \citet{Williams:2004}.  In this case, the range of dust temperatures considered here, i.e. 28 K, is broader than the range considered for the other disks, i.e. 20 K, because of the wider range of radial distances found for the unusually broad disk surrounding HD 107146 \citep{Ricci:2015}. The dependence of $\alpha_{\rm{Planck}}$ on the dust temperature is shown for each disk in Fig.~\ref{fig:dust_temperature}.

In Table~\ref{tbl:results}, the uncertainty on $q$ comes from the propagation of the uncertainties on $\alpha_{\rm{mm}}, \alpha_{\rm{Planck}}$ and $\beta_{s}$ using Eq.~\ref{eq_draine}.

\begin{deluxetable*}{lcccccccc}
\tablecaption{Results of the analysis\label{tbl:results}}
\tablehead{
  \colhead{Source}  &
  \colhead{Sub-mm/mm wavelength}     &
  \colhead{Flux density} &
  \colhead{Ref.$^a$} &
  \colhead{$\alpha_{\rm{mm}}$} &
  \colhead{$T_{\rm{dust}}$} &
  \colhead{Ref.$^a$} &
  \colhead{$\alpha_{\rm{Planck}}$} &
  \colhead{$q$} \\
  \colhead{}  &
  \colhead{[mm]}  &
  \colhead{[mJy]} &
  \colhead{} &
  \colhead{} &
  \colhead{[K]} &
  \colhead{} &
  \colhead{} &
  \colhead{} 
}
\startdata

q$^1$ Eri  & 0.35 & 179.9 $\pm$ 14.6 & 1 &  2.55 $\pm$ 0.07 & 49 $\pm$ 10 & 6 & 1.86 $^{+0.02}_{-0.04}$ & 3.38 $\pm$ 0.07 \vspace{1mm} \\
HD 107146  & 1.25 & 12.5 $\pm$ 1.3 & 2 & 2.55 $\pm$ 0.11 & 29$^{+23}_{-5}$ & 2 & 1.90$^{+0.05}_{-0.02}$ & 3.36$^{+0.07}_{-0.08}$ \vspace{1mm} \\
HD 181327  & 0.87 & 51.7 $\pm$ 6.2 & 3 & 2.86 $\pm$ 0.09 & 82 $\pm$ 10 & 7 & 1.96 $\pm$ 0.01
& 3.50 $\pm$ 0.07 \vspace{1mm} \\
HD 95086  & 0.35  & 112 $\pm$ 16 & 4 & 2.53 $\pm$ 0.10 & 55 $\pm$ 10 & 4 & 1.87$^{+0.02}_{-0.03}$ & 3.37 $\pm$ 0.07 \vspace{1mm} \\
$\beta$ Pic    & 0.87 & 60 $\pm$ 6$^b$ & 5 &  2.81 $\pm$ 0.10  & 85 $\pm$ 10 & 5  & 1.96 $\pm$ 0.01 & 3.47 $\pm$ 0.08 \\
\enddata
\tablenotetext{a}{References - 1) \citet{Eiroa:2011}, 2) \citet{Ricci:2015}, 3) \citet{Nilsson:2009}, 4) \citet{Su:2015}, 5) \citet{Dent:2014}, 6) \citet{Moro-Martin:2015}, 7) \citet{Mittal:2015}.}
\tablenotetext{b}{Observations of $\beta$ Pic are affected by spatial filtering (see text for more details).}

\end{deluxetable*}


\section{Discussion}
\label{sec:discussion}

In the previous Section we have estimated the slope $q$ of the solid size distribution for our five debris disks. Values range between 3.36 (HD 107146) and 3.50 (HD 181327) with uncertainties of 0.06 $-$ 0.08. 
Using the same analysis described here, \citet{Ricci:2012} obtained $q= 3.48 \pm 0.11$ in the case of Fomalhaut\footnote{Note that the uncertainty on the value reported here is slightly different than the one reported in \citet{Ricci:2012} because of a better treatment of the propagation of the uncertainties.}. The $q$-values derived for the different disks are consistent within the uncertainties, with a weighted mean $< q > = 3.42 \pm 0.03$. Note however that debris disks may in principle have different values for $q$. 

These values can be used to test predictions of models of collisional cascades. The typical benchmark model is the classical \citet{Dohnanyi:1969} prediction of $q = 3.51$ for a collisional cascade at the steady-state as due to the collective dynamical interaction of particles caused by inelastic collisions and fragmentation. This model assumes that neither the velocity dispersion of bodies nor their tensile strength is a function of the size of the bodies.

\citet{Pan:2012} have extended the \citet{Dohnanyi:1969} treatment by assuming that the velocity dispersion and tensile strength vary with the size of the bodies as power-laws. They found self-consistent steady-state solutions when viscous stirring balances velocity damping.
In these models, the predicted $q$-value depends on what holds a single body together (gravity or tensile strength) and on the mechanism which is driving the damping, i.e. catastrophic collisions, collisions with equal-sized bodies or with the smallest bodies in the system.
For solids with sizes $\sim 1-10$ mm, which are held together by tensile strength, the predicted $q$ values are $3.65 \leq q < 3.82$, $q = 4$, $4 \leq q < 4.33$ in the cases of catastrophic collisions, collisions with equal-sized bodies and with the smallest bodies, respectively \citep{Pan:2012}. 

\begin{figure*}[t!]
\centering
\includegraphics[scale=0.65]{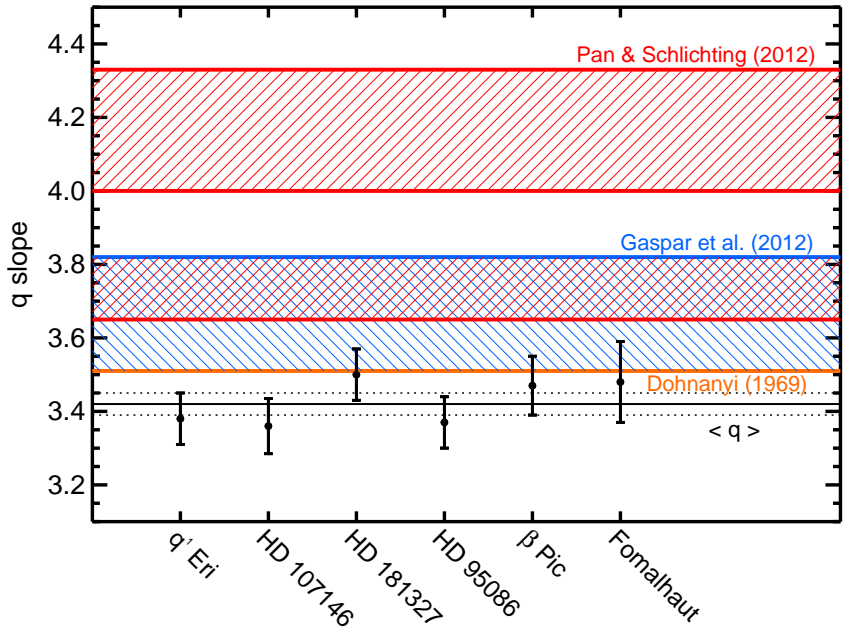}   
\vspace{-1mm}
\caption{Inferred $q$-values of the slope of the solid size distribution for our sample of debris disks together with model predictions. Data-points show the values  for the disks in our sample, including Fomalhaut from \citet{Ricci:2012}. The solid and dotted black horizontal lines show the weighted mean and uncertainty, i.e. $3.42 \pm 0.03$. The predictions from the different classes of models are indicated with the different colors. In particular, the horizontal orange line shows the single-valued classical \citep{Dohnanyi:1969} prediction, the blue horizontal band with oblique blue lines represent the \citet{Gaspar:2012} predictions, while the two red horizontal bands with oblique red lines show the values predicted by \citet[][see discussion in Section~\ref{sec:discussion}]{Pan:2012}.}
              \label{fig:q_disks}
\end{figure*}

\citet{Gaspar:2012} investigated the effects of model parameters on the evolution of the solid size distribution using numerical models of collision-dominated debris disks. Their calculations yield a quasi steady-state slope of $q = 3.65$, with a range between $q \approx 3.51 - 3.82$ when considering large variations for their model parameters.

When comparing these theoretical predictions with the results of our analysis, the classic prediction of the \citet{Dohnanyi:1969} collisional cascade lies at $\approx 3\sigma$ from the weighted mean of the $q$-values in our sample of disks (see Figure~\ref{fig:q_disks}).
The lowest values predicted by the \citet{Gaspar:2012} models are very similar to the \citet{Dohnanyi:1969} prediction, and therefore in line with the results for our debris disks, although these are obtained only with very large scaling factors for the strength curve, i.e. $S >> 10^{8}$ erg/g \citep{Benz:1999}.

Instead, the models by \citet{Pan:2012} predict values which are significantly larger than our results (Figure~\ref{fig:q_disks}). This may imply that the condition of exact balance between viscous stirring and some damping mechanism, which is the key assumption in those models, is not met in real disks.

A possible way to alleviate the apparent discrepancy of these findings with the model predictions is that part of the emission detected at 6.8 mm may not come from dust. If this was the case, our estimates would underestimate the true spectral index of the dust emission and therefore the $q$-value as well. For example, in the case of the AU Mic star, the emission seen by ALMA at 1.3~mm from the central regions of the debris disk can be explained by models of $\sim 10^6$ K plasma in an active stellar corona, which can also reproduce the strong X-ray flux from this source \citep{Cranmer:2013,MacGregor:2013}.  

For HD 107146, which shows the lowest inferred $q$-value among the debris disks in our sample, $\approx$ 65$\%$ of the measured flux at 6.8 mm would have to be produced by ionized gas to give a slope of the solid size distribution of $\approx 3.70$, which would be more in line with the predictions of the \citet{Pan:2012} and \citet{Gaspar:2012} models. However, the ALMA spatially resolved map for this disk at 1.25~mm does not show any hint for emission from the star location, as one would expect for emission from plasma in the stellar corona. The same can be argued for the spatially resolved ATCA maps at 6.8~mm for Fomalhaut \citep{Ricci:2012} and $\beta$ Pic (Figure~\ref{fig:betapic}), which do not show any strong emission from the stellar location.

For the disks which have not been yet spatially resolved at millimeter wavelengths, observations with higher angular resolution are needed to investigate possible ionized gas emission from the stellar corona.
Also, observations at even longer wavelengths in the radio have the ability to disentangle the contribution of dust and ionized gas from the radio spectrum.

\section{Summary}
\label{sec:summary}

We have presented new ATCA observations for the continuum emission at 6.8 mm of five debris disks, i.e. $q^1$ Eri, HD 107146, HD 181327, HD 95086, $\beta$ Pic. By combining these new detections with those at shorter sub-mm/mm wavelengths we have measured the sub-mm/mm spectral index for these sources. 

At these long wavelengths the dust emission is close to the Rayleigh-Jeans approximation, and we used previous estimates for the dust temperature in each of these systems to quantify the deviation from the Rayleigh-Jeans regime. We derived the spectral index $\beta$ of the dust emissivity ($\kappa_{\nu} \propto \nu^{\beta}$), and used a relation derived by \citet{Draine:2006} to infer the slope $q$ of the solid size distribution ($n(a) \propto a^{-q}$).

The slopes we derived range between $\approx$ 3.36 and 3.50 with values consistent within the errorbars and a weighted mean $< q > = 3.42 \pm 0.03$. We compared these findings with the predictions of models of collisional cascades in debris disks.  
Models by \citet{Pan:2012}, which assume a balance between viscous stirring and some damping mechanism, overpredict these estimates significantly. Numerical models by \citet{Gaspar:2012} can predict values as low as $\approx 3.5$, but only by invoking very large scaling factors for the strength curve ($S >> 10^{8}$~erg/g). With a prediction of $3.51$, the classical models of \citet{Dohnanyi:1969} appear to be the closest to reproduce our results. This may indicate that the variation of the solid velocity dispersions and/or collision parameters are weaker than previously thought.


\acknowledgments


\end{document}